\documentclass[aps,showpacs,preprintnumbers]{revtex4}
\usepackage{amsfonts}
\usepackage{bm}
\usepackage[dvips]{graphicx}

\begin{document}

\title{Analytic result for the one-loop massless triangle Feynman diagram}
\author{A.T.Suzuki}
\affiliation{Instituto de F\'{\i}sica Te\'orica-UNESP,\\
Rua Pamplona 145 -- 01405-900 S\~ao Paulo, SP -- Brazil}
\author{}
\affiliation{}
\author{}
\affiliation{}
\date{\today }

\begin{abstract}
Different mathematical methods have been applied to obtain the analytic result for the massless triangle Feynman diagram yielding a sum of four linearly independent hypergeometric functions $F_4$. In this paper I work out the diagram and show that that result, though mathematically sound, is not physically correct, because it misses a fundamental physical constraint imposed by the conservation of momentum, which should reduce by one the total number of linearly independent (l.i.) functions $F_4$ in the overall solution. Taking into account that the momenta flowing along the three legs of the diagram are constrained by momentum conservation, the number of overall l.i. functions that enter the most general solution must reduce accordingly.

To determine the exact structure and content of the analytic solution for the three-point function, I use the analogy that exists between Feynman diagrams and electric circuit networks, in which the electric current flowing in the network plays the role of the momentum flowing in the lines of a Feynman diagram. This analogy is employed to define exactly which three out of the four hypergeometric functions are relevant to the analytic solution for the Feynman diagram. The analogy is built based on the equivalence between electric resistance circuit networks of type ``Y'' and ``Delta'' in which flows a conserved current. The equivalence is established via the theorem of minimum energy dissipation within circuits having these structures. 
\end{abstract}

\pacs{11.10.Gh,11.15.Bt,11.55.Bq, 12.38.Bx, 14.70.Dj}
\maketitle







\section{Introduction}

Massless triangle Feynman diagrams occur in field theoretical perturbative calculations, being one of the primary divergent (one-particle irreducible) graphs for the non-Abelian gauge theories such as Yang-Mills fields. This makes this kind of perturbative calculation the more interesting and necessary once we want to check the higher order quantum corrections for many physical processes of interest. Two decades ago, Boos and Davydychev \cite {BoosDavy} were the first to obtain the analytic formulae for the one-loop massless vertex diagram using the Mellin-Barnes complex contour integral representation for the propagators. Later on, Suzuki, Santos and Schmidt \cite {SSS} reproduced the same result making use of the negative dimensional integration method (NDIM) technique of Halliday and Ricotta \cite{HR} for Feynman integrals, where propagators are initially taken to be finite polynomials in the integrand, and once the result is obtained, analytically continued to the realm of positive dimensions.   

These two widely different methods of calculation yielding the same answer lends to the analytic result obtained a degree of certainty as far as the mathematical soundness is concerned. However, here I argue that mathematical soundness only is not enough when we are interested in processes of physical content. 

Neither the Mellin-Barnes formulation of Boos and Davydychev nor the NDIM technique employed by Suzuki {\sl et al} to evaluate the massless triangle graph take into account the basic physical constraint imposed on such a diagram due to the momentum conservation flowing in the three legs. Momentum (that is, energy and three-momentum) conservation is one of the basic, fundamental tenets of modern natural sciences, and plays a key role in determining the final form for the analytic result for the massles one-loop vertex correction. 

The system of coupled, simultaneous partial differential equations:
\begin{eqnarray}
x(1-x)\frac{\partial^2 Z}{\partial x^2}-y^2\frac{\partial^2 Z}{\partial y^2}-2xy\frac{\partial^2 Z}{\partial x \partial y}
+ [\gamma-(\alpha+\beta+1)x]\frac{\partial Z}{\partial x} - (\alpha+\beta+1)y\frac{\partial Z}{\partial y}-\alpha \beta Z =  0 \nonumber \\
y(1-y)\frac{\partial^2 Z}{\partial y^2}-x^2\frac{\partial^2 Z}{\partial x^2}-2xy\frac{\partial^2 Z}{\partial x \partial y}
+ [\gamma\hspace{.05cm}'-(\alpha+\beta+1)y]\frac{\partial Z}{\partial y} - (\alpha+\beta+1)x\frac{\partial Z}{\partial x}-\alpha \beta Z  = 0 
\end{eqnarray}
has a solution which consists of a linear combination of four hypergeometric functions of type $F_4$, as follows \cite{Appell}:
\begin{eqnarray}
Z & = & {\bf A}\,F_4(\alpha, \, \beta, \, \gamma, \, \gamma\hspace{.05cm}';\, x, \, y)\label{A} \\
&+& {\bf B}\,x^{1-\gamma}\,F_4(\alpha+1-\gamma, \, \beta+1-\gamma, \, 2-\gamma, \, \gamma\hspace{.05cm}'; \, x, \, y)\label{B} \\
&+& {\bf C}\,y^{1-\gamma\hspace{.05cm}'}\,F_4(\alpha+1-\gamma\hspace{.05cm}', \, \beta+1-\gamma\hspace{.05cm}', \, \gamma, \, 2-\gamma\hspace{.05cm}'; \, x, \, y)\label{C} \\
&+& {\bf D}\,x^{1-\gamma}y^{1-\gamma\hspace{.05cm}'}\,F_4(\alpha+2-\gamma-\gamma\hspace{.05cm}', \, \beta+2-\gamma-\gamma\hspace{.05cm}', \, 2-\gamma, \, 2-\gamma\hspace{.05cm}'; \,x, \, y) \label{D1}
\end{eqnarray}
where it is implicit that $x$ and $y$ are independent variables. 

Now the analytic solution for the massless triangle Feynman diagram given by Boos and Davydychev \cite{BoosDavy} and Suzuki {\sl et al} \cite {SSS} has exactly the same structure as the above solution for the simultaneous partial differential equation. However, the variables $x$ and $y$ for the triangle Feynman diagram are given by ratios of momentum squared, such as $x=\displaystyle\frac{r^2}{p^{2}}$ and $y=\displaystyle\frac{q^2}{p^{2}}$ where $r=q-p$, so that the variables $r,\:q$ and $p$ are not independent, but constrained by momentum conservation. Therefore, the solution for the Feynman diagram cannot be a linear combination of four linearly independent hypergeometric functions $F_4$, but a linear combination of {\sl three} linearly independent hypergeometric solutions $F_4$. 

One of the simplest examples where the four-term hypergeometric combination above mentioned does not reproduce the correct result is when it is embedded in a higher two-loop order calculation. Consider, for example, the diagrams of figure 1. 

\begin{figure}[h]
\centering
\includegraphics[height=1.6483in,width=4.7899in]{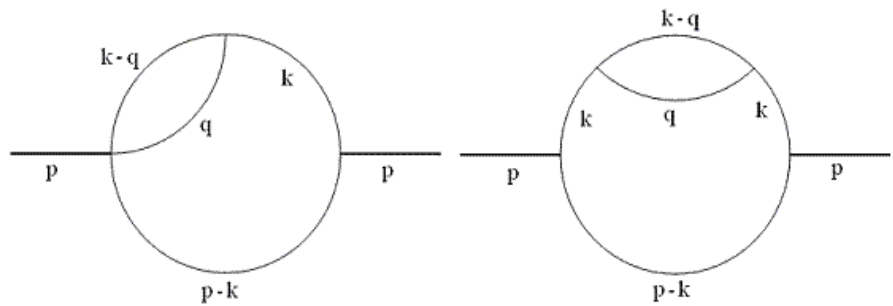}
\caption{Two-point two-loop Feynman diagrams: Fying-saucer side-view and front-view.}
\label{Figure 1:}
\end{figure}

For neither of the diagrams the four-term solution for the triangle when embedded into the two-loop graph yields the correct answer. The right result can only be achieved when we use a three-term triangle solution embedded into the two-loop structure \cite{SSB}.

One could use the analytic continuation formula for the $F_4$ function \cite{Appell} in order to reduce the number of independent functions to three; however, one has at least two possibilities to do this: Either combining (\ref{A}) and (\ref{C}) or combining (\ref{B}) and (\ref{D1}), so that, for example, for the former case
\begin{eqnarray}
\label{anacont}
{\bf A}\,F_4(\alpha,\,\beta,\,\gamma,\,\gamma\,';\,x,\,y)
            &+& {\bf C}\,F_4(\beta+1-\gamma\hspace{.05cm}', \, \alpha+1-\gamma\,', \, \gamma, \, 2-\gamma\hspace{.05cm}'; \, x, \, y)\nonumber \\
          &=& {\bf \Omega}\, F_4\left(\alpha, \, \alpha+1-\gamma\,', \, \gamma, \, \alpha+1-\beta; \, \frac{x}{y}, \, \frac{1}{y}\right)
\end{eqnarray}

Therefore, which three of those $F_4$'s should be considered cannot be determined by the momentum conservation only; it needs another physical input to this end.  

In order to determine the exact structure and content of the analytic result for the triangle diagram I employ the analogy that exists between Feynman diagrams and electric circuit networks. The conclusion of the matter can be summarized as follows: since there is a constraint (call it either a initial condition or a boundary condition) to the momenta flowing in the legs of a triangle diagram, it means that the overall analytic answer obtained via Mellin-Barnes or NDIM technique, which comes as a sum of four linearly independent hypergeometric functions of type $F_4$, in fact must contain only three linearly independent ones. Which three of these should be is determined by the momentum conservation flowing through the legs of the diagram, {\sl and}, by the circuit analogy, by the equivalence between the ``Y-type'' and ``$\Delta$-type'' resistance networks through which {\sl conserved} electric current flows.          
  
\section{Electric circuitry and Feynman diagrams}

Although known for a long time and mentioned sometimes in the literature of field theory, the analogy between electric circuit networks and Feynman diagrams, more often than not stays as a mere curiosity or at a diagrammatic level in which the drawing is only a representation that helps us out in the ``visualization'' of a given physical process. There is, however, a few exceptions to this. The earliest one that I know of is by Mathews \cite {Mathews} as early as 1958 where he deals with singularities of Green's functions and another one by Wu \cite {Wu} in 1961, where he gave an algebraic proof for several properties of normal thresholds (singularities of the scattering amplitudes) in perturbation theory. 

What I am going to do here is just to state the well-known result for the equivalence between ``$Y$''- and ``$\Delta$-circuits'' which is used, for example, in analysing the ``Kelvin bridge'' in relation to the more commom ``Wheatstone brigde''. The equivalence I consider is for the network of resistors, where energy is dissipated via Joule effect. Then with this in hands I argue which three out of the overall four hypergeometric functions $F_4$ present in the one-loop vertex obtained via Mellin-Barnes or via NDIM should be considered as physically meaningful.  

Consider then the ``$Y$''- and ``$\Delta$-circuits'' of figure 2. 

\begin{figure}[h]
\centering
\includegraphics[height=1.6483in,width=3.7899in]{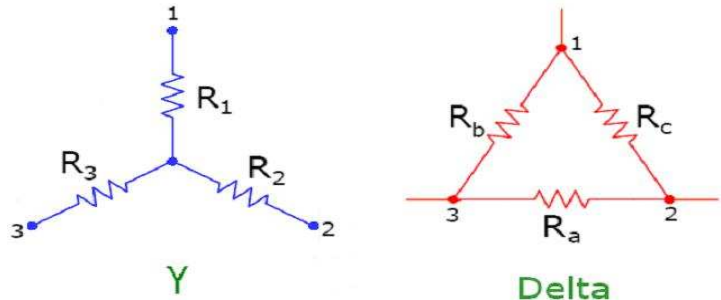}
\caption{``Y-type'' and ``Delta-type'' resistor networks.}
\label{Figure 2:}
\end{figure}

These two network of resistors are related to each other; for example, the ``Delta-circuit'' can be made equivalent to the ``Y-circuit'' when resistors obey the following equations \cite {Farago}:
\begin{eqnarray}
R_a&=&R_2+R_3+\frac{R_2 R_3}{R_1} \nonumber\\
R_b&=&R_1+R_3+\frac{R_1 R_3}{R_2} \label{Ra}\\
R_c&=&R_1+R_2+\frac{R_1 R_2}{R_3} \nonumber
\end{eqnarray}

In electric circuits a general theorem guarantees that the current density in a conductor distributes itself in such a way that the generation of heat is a minimum \cite{Smythe}. Strictly speaking, since I consider the ``Y''- and ``Delta-networks'' by themselves and not embedded within an electrical circuit, I cannot consider the minimum of power generation, but still I can consider the equivalence between the power generated in one of them --- say ``Y-circuit'' --- as compared to the power generated in the other one, the ``Delta-circuit''.     

Labelling the currents that flow through resistors $R_1$, $R_2$ and $R_3$ of the ``Y-circuit'' by $p$, $q$ and $r$, respectively, by reason of current conservation we have that, for example \footnote[1]{Of course, we can take either two of those as independent ones, such that the dependent current can be expressed also as either $q = p+r$ or $p = q-r$}
\begin{equation}
r = q - p
\end{equation}

Each of these currents will generate heat according to Joule's law and the corresponding power will be given by Ohm's law. So in each leg, the product of the resistance times the square of the current will give the power generated by the current flowing through it. The overall power generated in the ``Y-network'' is therefore: 
\begin{equation}
p^2R_1+q^2R_2+r^2R_3=P_{\rm total}
\end{equation}
From this we have the following system of equations:
\begin{eqnarray}
R_1+\lambda R_2+\mu R_3&=&\frac{1}{p^2}P_{\rm total}\equiv {\bf R}_{\rm p} \label{D}\\
\frac{1}{\lambda}R_1+R_2+\nu R_3&=&\frac{1}{q^2}P_{\rm total}=\frac{1}{\lambda}{\bf R}_{\rm p}\label{E}\\
\frac{1}{\mu}R_1+\frac{1}{\nu}R_2+R_3&=&\frac{1}{r^2}P_{\rm total}=\frac{1}{\mu}{\bf R}_{\rm p} \label{F}
\end{eqnarray}
where for convenience I have defined  $\lambda \equiv \displaystyle\frac{q^2}{p^2}$, $\mu \equiv \displaystyle\frac{r^2}{p^2}$, $\nu \equiv \displaystyle\frac{r^2}{q^2}$ and ${\bf R}_{\rm p} \equiv \displaystyle\frac{1}{p^2}P_{\rm total}$. 

Multiplying the last equation by $\mu$ and comparing with (\ref{D}),
\begin{eqnarray}
R_1+\lambda R_2+\mu R_3&=&\frac{1}{p^2}P_{\rm total}\equiv {\bf R}_{\rm p} \label{G}\\
R_1+\frac{\mu}{\nu}R_2+\mu R_3&=&\frac{\mu}{r^2} P_{\rm total}={\bf R}_{\rm p} \label{H}
\end{eqnarray}
it follows that the coefficient of $R_2$ is $\lambda = \displaystyle{\mu}{\nu}$.

Multiplying (\ref{D}) by $\displaystyle\frac{1}{\lambda}$ and comparing with (\ref{E}),
\begin{eqnarray}
\frac{1}{\lambda}R_1+R_2+\frac{\mu}{\lambda} R_3&=&\frac{1}{\lambda p^{2}}P_{\rm total}\equiv \frac{1}{\lambda}{\bf R}_{\rm p} \label{I}\\
\frac{1}{\lambda}R_1+R_2+\nu R_3&=&\frac{1}{q^{2}}P_{\rm total}=\frac{1}{\lambda}{\bf R}_{\rm p}\label{J}
\end{eqnarray}
it follows that the coefficient of $R_3$ is $\nu=\displaystyle\frac{\mu}{\lambda}$.

Finally, multiplying (\ref{F}) by $\lambda \nu$ and comparing with (\ref{D}),
\begin{eqnarray}
R_1+\lambda R_2+\mu R_3&=&\frac{1}{p^{2}}P_{\rm total}\equiv {\bf R}_{\rm p} \label{K}\\
\frac{\lambda \nu}{ \mu}R_1+\lambda R_2+\mu R_3&=&\frac{\mu}{ r^{2}}P_{\rm total}={\bf R}_{\rm p} \label{L}
\end{eqnarray}
it follows that the coefficient of $R_1$ is $1=\displaystyle\frac{\lambda \nu}{ \mu}$.

On the other hand, in the ``Delta-network'', let $I_a$ label the current flowing through resistance $R_a$, so that the power generated in this resistor is, using (\ref{Ra})
\begin{equation}
P_{\rm a} = I_{a}^2\,R_a = I_{a}^2\,\left \{\frac{R_2\,R_3}{R_1}+R_2+R_3\right \}
\end{equation}

Now, as the resistances follow the equivalence defined by (\ref{Ra}), the current $I_a$ that flows through $R_a$ must be such that:
\begin{equation}
I_a^2R_a \equiv p_a^2\,R_{1,a}+q_a^2\,R_{2,a}+r_a^2\,R_{3,a}
\end{equation}
where
\begin{eqnarray}
R_{1,a}&=&\frac{R_2\,R_3}{R_1}\nonumber\\
R_{2,a}&=&R_2\\
R_{3,a}&=&R_3 \nonumber
\end{eqnarray}
and
\begin{eqnarray}
p_a^2 &=&\frac{\lambda\,\nu}{\mu}\,p^2=p^2 \nonumber\\
q_a^2&=&\lambda\,q^2\\
r_a^2&=&\nu\,r^2
\end{eqnarray}

Therefore, the solution for the one-loop triangle Feynman diagram must be such that it contains three terms which are proportional to ``currents'' $p^2$, $\lambda\,q^2$ and $\nu\,r^2$. This solution is exactly achieved when I combine (\ref{B}) and (\ref{D1}) in a similar way as was done in (\ref{anacont}). Combination of (\ref{A}) and (\ref{C}) as in (\ref{anacont}) leaves the term (\ref{D1}) in the solution, which is proportional to $\mu \lambda\,p^{2}$, and therefore not suitable since it violates the equivalence above demonstrated.  

Explicit analytic solution for the one-loop massless triangle Feynman diagram reads therefore
\begin{eqnarray}
\Sigma_{\rm one-loop}(p,q,r)&=& (p^2)^{\sigma}\,{\bf \Gamma_p}\,F_4(\alpha_p,\,\beta_p,\,\gamma_p,\,\gamma\,'_p;\,\mu,\,\lambda)\nonumber\\
&+&\lambda\,(q^2)^{\sigma}\,{\bf \Gamma_q}\,F_4(\alpha_q,\,\beta_q,\,\gamma_q,\,\gamma\,'_q;\,\mu,\,\lambda)\\
&+&\nu\,(r^2)^{\sigma}\,{\bf \Gamma_r}\,F_4\left(\alpha_r,\,\beta_r,\,\gamma_r,\,\gamma\,'_r;\,\nu,\,\frac{1}{\lambda}\right)\nonumber
\end{eqnarray}
where $\sigma = D/2-3$ with $D$ being the dimensional regularization parameter. One notes that the variables for the last $F_4$, though related to those that appear in the first and second $F_4$'s, differ from them. 

\section{Conclusion}

Using the basic physical principle of momentum conservation I deduced the correct analytic result for the one-loop massless triangle Feynman diagram in terms of three linearly independent hypergeometric functions $F_4(\alpha,\,\beta,\,\gamma,\,\gamma\,';\,\mu,\,\lambda)$. The reason why there must be only three linearly independent functions in the solution is due to the fact that the variables $\mu$ and $\lambda$ are such that implicit in them is a momentum conservation constraint that must be taken into account properly. The set of which three linearly independent functions $F_4$ should be is given by another physical input, deduced in analogy to the equivalence between ``Y''- and ``Delta-network'' of electrical resistance circuits. With the conservation of momentum correctly taken into account, the final result for the mentioned Feynman diagram is the physically correct and relevant analytic solution to the problem, having only a combination of three linearly independent $F_4$ functions in it.       

\vspace{1cm}
{\tt Acknowledgments} Work dedicated to Mitiko, my beloved wife, and to our children: Tamie, on her 15th birthday, Tise on her 13th birthday and Timothy on his 7th birthday.


\vspace{1cm}

\begin{references}

\bibitem{BoosDavy}  E.E.Boos and A.I.Davydychev, {\sl Vestnik MGU} {\bf 28}, (1987) 8. Cited by V.A.Smirnov in his book ''Renormalization and Asymptotic Expansions'', page 124. Volume 14 of the series Progress in Physics, Birkh\"auser Verlag, Basel, Boston, Berlim (1991) 

\bibitem{SSS}  A.T.Suzuki, E.S.Santos and A.G.M.Schmidt, {\sl Eur.Phys.J.} {\bf C26} (2002) 125-137.

\bibitem{HR} I.G.Halliday and R.M.Ricotta, {\sl Phys. Lett.} {\bf B193} (1987) 241.

\bibitem{Appell} P.Appell, and J.Kamp\'e de F\'eriet, {\sl Fonctions Hyperg\'eom\'etriques et Hypersph\'eriques --- Polynomes d'Hermite}, Gauthier-Villars et Cie. Editeurs, Paris (1926) 52 

\bibitem{SSB} A.T.Suzuki, A.G.M.Schmidt, and J.D.Bolzan, arXiv hep-ph-0808.1496v1

\bibitem{Mathews}  J.Mathews, {\sl Phys.Rev.} {\bf 113} nb. 1 (1958) 381.

\bibitem{Wu} T.T.Wu, {\sl Phys. Rev.} {\bf 123} nb. 2 (1961) 689-691

\bibitem{Farago} P.S.Farago, {\sl An Introduction to Linear Network Analysis}, The English Universities Press Ltd., London (1961) pages 19-21  

\bibitem{Smythe} W.R.Smythe, {\sl Static and Dynamic Electricity}, McGraw Hill Book Co. Inc., New York-Toronto-London (1950), page 233
 
\end{references}
\end{document}